\documentclass[prd,aps,showpacs,tightenlines]{revtex4}  %%preprint,
\usepackage{mathrsfs}
\usepackage{amsmath}
\usepackage{amssymb}
\usepackage{epsfig}
\usepackage{color}
\usepackage{subfigure}
\begin{document}

\title{The S-wave resonance contributions in the  $B^0_s$ decays into $ \psi(2S,3S)$ plus  pion pair}
\author{Zhou Rui$^1$}            \email{jindui1127@126.com}
\author{Ya Li$^2$}                   \email{liyakelly@163.com}
\author{Wen-Fei Wang$^3$}   \email{wfwang@sxu.edu.cn}

\affiliation{$^1$College of Sciences, North China University of Science and Technology,
                          Tangshan 063009,  China}
 \affiliation{$^2$Department of Physics and Institute of Theoretical Physics,
                          Nanjing Normal University, Nanjing 210023, Jiangsu ,   China}
 \affiliation{$^3$Institute of Theoretical Physics, Shanxi University, Taiyuan 030006, Shanxi, China}

\date{\today}

\begin{abstract}
The three-body decays $B^0_s \rightarrow \psi(2S,3S) \pi^+ \pi^-$ are studied  based on the perturbative QCD approach.
With the help of the nonperturbative two-pion distribution amplitudes, the analysis is simplified into the quasi-two-body
processes. Besides the traditional factorizable and nonfactorizable  diagrams at the leading order,
the  next-to-leading order vertex corrections are also included to  cancel the scale dependence.
The $f_0(980)$,  $f_0(1500)$ resonance contributions as well as the  nonresonant contributions  are taken into account
using the presently known $\pi\pi$  time-like scalar form factor for the $s\bar{s}$ component.
It is found that the predicted  $B^0_s \rightarrow \psi(2S) \pi^+ \pi^-$ decay spectra in the pion pair invariant mass shows
a similar behavior as the experiment. The calculated S-wave contributions to the branching ratio of
$B^0_s \rightarrow \psi(2S) \pi^+ \pi^-$ is $6.0\times 10^{-5}$, which is in agreement with the LHCb data
$\mathcal {B}(B^0_s \rightarrow \psi(2S) \pi^+ \pi^-)=(7.2\pm 1.2)\times 10^{-5} $   within errors.
The estimate  of  $\mathcal {B}(B^0_s \rightarrow \psi(3S) \pi^+ \pi^-)$ can reach
the  order of $10^{-5}$, pending the corresponding measurements.
\end{abstract}


\maketitle

\section{Introduction}

%{\color{red} [All the CP should be $CP$]}

%The three-body hadronic $B$ meson decays play an important role in the hunting of the physics beyond Standard Model.
%in particular to determine the Cabibbo-Kobayashi-Maskawa (CKM) couplings and to study CP violation and mixing. Furthermore,
The decays of the neutral $B$ mesons to the charmonium state plus  light meson pair have attracted lots of attentions recently.
These decay  processes  could be used to study the spectroscopy in various meson pair systems, and they also have significant roles
in understanding of the substructure of different resonant states.
Up to now, several experimental collaborations, like LHCb and BaBar, have  measured  the decays $B_s^0\rightarrow J/\psi \pi^+\pi^-$ \cite{prd86052006,prd89092006}, $B_s^0\rightarrow J/\psi K^+ K^-$ \cite{prd87072004,prl114041801},
$B^0\rightarrow J/\psi \pi^+\pi^-$\cite{prl90091801,prd87052001,prd90012003,plb74238}, $B^0\rightarrow J/\psi K^+ K^-$ \cite{prd88072005},
$B^0_s\rightarrow J/\psi \phi\phi$ \cite{jhep03040},
$B^0_s\rightarrow \psi(2S) K^+\pi^-$ \cite{plb747484}.

On the theoretical side, the studies of $B$ mesons decay into three hadronic final states have been
performed using different approaches~\cite{plb56490,prd72094031,plb727136,plb726337,prd89074043,plb728579,prd91014029,
HYCPRD1,HYCPRD2,HYCPRD3,HYCPRD4,prd87076007,epjc75536,prd89094007,pma58031001,npb899247,BEI1,BEI2,prd81094033,prd89094013,prd92054010,
jhep02009,epjc75609,ps88035101,jhep09074} in recent years.
The perturbative QCD approach (PQCD) \cite{pqcd1,pqcd2}
%is one of the recently developed theoretical frameworks  to deal with the nonleptonic, semileptonic, and,
has some unique features that are particularly suitable for dealing with the three-body $B$ meson decays.
%For the three-body $B$ meson decays, in the  perturbative QCD approach (PQCD)~\cite{pqcd1,pqcd2},
First, by introducing two-meson distribution amplitudes ~\cite{fp42101,prl811782,prd62073014,epjc26261,npb555231,sjnp38289,tmp691109},  the analysis of three-body hadronic $B$
meson decays is simplified into the one for two-body decays.
As pointed out in ~\cite{plb561258,prd70054006}, it is not practical to make a direct calculation for the three-body processes due to the enormous number of
 diagrams which contain two gluon exchange at lowest order. Besides, its contribution is not important because the region with the two gluons being hard
  simultaneously is power suppressed.
%can be reduced to a process via quasi-two body decays containing a resonance state.
The dominant contributions come from the kinematic region where the two light mesons move almost parallelly for producing a resonance with an invariant mass below $O(\bar{\Lambda}M_B)$ ($\bar{\Lambda}$ being the heavy-meson and heavy-quark mass difference),
which can be catched by  this new nonperturbative inputs.
Second, the three-body decays of $B$ meson
receive both resonant and nonresonant contributions, it is difficult to separate the two parts  clearly ~\cite{npb899247}. After absorbing the nonperturbative  dynamics associated with the meson pair into the complex time-like form factors in the  two-meson distribution amplitudes,
%the universal two-meson distribution amplitudes,
 both resonant and nonresonant contributions~\cite{plb561258,prd70054006} are
included in the PQCD approach. Third, in the above dominant contribution region, the end-point singularities are smeared by
the two-meson invariant mass, which suggest that the PQCD approach has a good predictive power without any arbitrary cutoffs. Finally,
different from the two-body  cases, some possible  final state interactions (FSIs) may be significant in the three-body decays~\cite{prd89094013, fsi2}.
According to \cite{fsi2}, there are two distinct FSIs mechanisms. One is the interactions between the meson pair in the resonant region associated with
 various intermediate states. The other is the  rescattering between the third particle  and the pair of mesons. In our opinion, the former can be factorized into two-meson distribution amplitudes while the latter is ignored in the quasi-two-body approximation.
%The nonperturbative  dynamics associated with the meson pair  can be absorbed into the universal two-meson distribution amplitudes \cite{fp42101,prl811782,prd62073014,epjc26261,npb555231,sjnp38289,tmp691109}
%which contain  both resonant and nonresonant contributions~\cite{plb561258,prd70054006}.
Over past few years, the braching ratios and direct $CP$ asymmetries of the three-body decays such as
 $B^{\pm}\rightarrow \pi^{\pm}(K^{\pm})\pi^+\pi^- $ \cite{prd89074031}, $B^0_{(s)}\rightarrow J/\psi \pi^+ \pi^-$ \cite{prd91094024},
 $B^0_{(s)}\rightarrow \eta_c \pi^+\pi^-$ \cite{150906117},   $B^+_c\rightarrow D^+_{(s)}\pi^+\pi^-$ \cite{prd94034040}, $B_{(s)}\rightarrow (D_{(s)}, \bar{D}_{(s)})\pi^+\pi^-$ \cite{161108786}, $B_{(s)}\rightarrow P\pi^+\pi^-$ \cite{161205934} and $B^0_{(s)}\rightarrow \eta_c (2S)\pi^+\pi^-$ \cite{170101844}
have been studied systematically in the PQCD approach.
In a recent work, the P-wave resonance contributions have been calculated in the
$B\rightarrow K \pi\pi$ decays \cite{160904614}.

The three-body hadronic $B$ meson decays with the radially excited charmonium mesons in the final state have not received
much attentions in the literature. As the LHC gathering more and more data, the processes of the $B$
mesons decays including excited charmonium states must have much possibilities to be found.
Recently, the LHCb Collaboration have measured  the three-body decay channels of $B^0_{(s)}\rightarrow \psi(2S) \pi^+ \pi^-$
with $pp$ collision data collected at $\sqrt{s}=7$ TeV recently~\cite{npb871403}.
In the quark model, $\psi(2S)$ ($\psi(3S)$)  is the first  (second) radially excited vector charmonium with the radial quantum number $n=2(3)$ and
the orbital angular momentum $l=0$. Both   $\psi(2S)$ and $\psi(3S)$ were  observed by the processes of the  $e^+e^-$ annihilation into  hadronic \cite{prl331453,dasp}. The  properties for the two high excited  charmonium were updated in PDG 2016 \cite{pdg}; they are listed in Table \ref{tab:pro}.
\begin{table}\label{tab:pro}
\caption{The properties of $\psi(2S)$ and $\psi(3S)$ mesons.}
\begin{tabular}[t]{lcccc}
\hline\hline
%\toprule[2pt]
Mesons          &$I^G$   &$J^{PC}$              &Mass (MeV) & Width (keV)
 \\ \hline
$\psi(2S)$  & $0^-$ & $1^{--}$&$3686.097\pm 0.025$& $296\pm 8$\\
$\psi(3S)$  & $0^-$ & $1^{--}$&$4039\pm 1$ &$800\pm 100$\\
\hline\hline
\end{tabular}
\end{table}
%Now the PDG have given the mass and width of
%It is necessary to confirm the above experimental results in various theoretical frameworks.
%{\color{red} [Some introductions about $\psi(2s,3s)$ needed here]}

\begin{figure}[tbp]
%\vspace{-1cm}
\centerline{\epsfxsize=9cm \epsffile{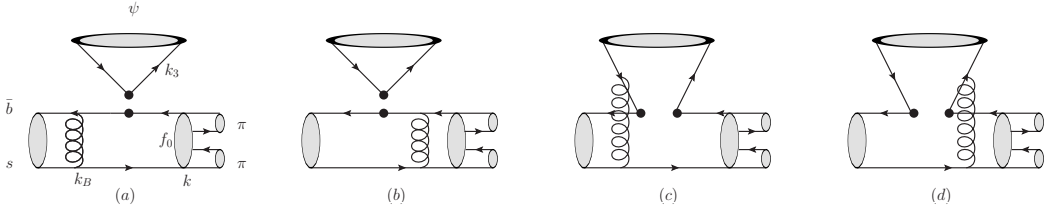}}
\vspace{-8cm}
%\vspace{-6cm}
\caption{The leading-order Feynman diagrams for the quasi-two-body decays
              $B^0_s\rightarrow \psi(2S,3S)f_0\rightarrow \psi(2S,3S)\pi^+\pi^-$, where $f_0$ stands for the S-wave intermediate state.}
\label{fig:femy}
\end{figure}

We have previously studied the semileptonic and two-body nonleptonic decays of the $B_c$ meson to
radially excited charmonium mesons in the PQCD approach by using the harmonicosillator wave functions for
the charmonium states~\cite{160208918,epjc75293}.
In the present work, we will extend our analysis to the three-body decays
$B^0_s\rightarrow \psi(2S,3S)\pi^+\pi^-$ to provide an order of magnitude estimation.
By introducing the two-pion distribution amplitudes, the S-wave contributions, which are the main contributions
in the three-body decays $B^0_s\rightarrow \psi(2S,3S)\pi^+\pi^-$~\cite{npb871403},
could be described by the quasi-two-body processes $B^0_s\rightarrow \psi(2S,3S)f_0\rightarrow \psi(2S,3S)\pi^+\pi^-$
containing the S-wave resonant states $f_0$,  $f_0(980)$ and $f_0(1500)$ as two examples.
Following the steps in Ref. \cite{prd91094024},
the decay amplitude $\mathcal{A}(B^0_s\rightarrow \psi(2S,3S)\pi^+\pi^-)$  can be written as the convolution
%the pQCD factorization formula for the  quasi-two-body decays
\begin{eqnarray}
\mathcal{A}(B^0_s\rightarrow \psi(2S,3S)\pi^+\pi^-)=H \otimes\phi_B\otimes \phi_{\psi} \otimes \phi^S_{\pi\pi},
\end{eqnarray}
where the hard function $H$ could be treated by PQCD including both  factorizable and
nonfactorizable contributions in the expansion at the leading order in $\alpha_s$   (single gluon
exchange as depicted in Fig.\ref{fig:femy}).
The hadron wave functions $\phi_B, \phi_{\psi}$ and $\phi^S_{\pi\pi}$ absorb the nonperturbative dynamics, and they can be
extracted from experimental data or other nonperturbative methods.

Following the introduction, in Sect. \ref{sec:framework},
we shall define kinematics and describe the wave functions of  the initial and final states,
then we will briefly review the  related  theoretical formulas.
In Sect. \ref{sec:results}, we will calculate the $B^0_s\rightarrow \psi(2S,3S)\pi^+\pi^-$
decays in the PQCD approach with discussions.  Finally we will close this paper with a conclusion.

\section{framework}\label{sec:framework}

It is convenient to work in the rest frame of the $B_s^0$ meson. % and in the light cone coordinate.
Its momentum $p_B$, along with the charmonium meson momentum $p_3$, the pion pair momentum $p$ and
other quark momenta $k_i$ in each meson, is chosen as \cite{prd91094024}
\begin{eqnarray}
 p_B&=&\frac{M_B}{\sqrt{2}}(1,1,\textbf{0}_{\rm T}),\quad p_3=\frac{M_B}{\sqrt{2}}(r^2,1-\eta,\textbf{0}_{\rm T}),\quad  p=\frac{M_B}{\sqrt{2}}(1-r^2,\eta,\textbf{0}_{\rm T}),\nonumber\\
  k_B&=&(0,\frac{M_B}{\sqrt{2}}x_B,\textbf{k}_{\rm BT}),\quad k_3=(\frac{M_B}{\sqrt{2}}r^2x_3,\frac{M}{\sqrt{2}}(1-\eta)x_3,\textbf{k}_{3\rm T}),\quad  k=(\frac{M_B}{\sqrt{2}}z(1-r^2),0,\textbf{k}_{\rm T}),
\end{eqnarray}
with the ratio $r=m/M_B$ and $m (M_B)$ is the mass of the charmonium ($B^0_s$) meson.
The factor $\eta=\omega^2/(M_B^2-m^2)$ with the invariant mass squared $\omega^2=p^2$
for the pion pair. The $k_{iT}$, $x_i$ represent the transverse momentum and longitudinal
momentum fraction of the quark inside the meson, respectively.
If we choose the $\zeta=p_1^+/p^+$ as the $\pi^+$ meson momentum fraction,  the two pion
momenta $p_{1,2}$ can be written as
\begin{eqnarray}
 p_1=\frac{M_B}{\sqrt{2}}((1-r^2)\zeta, \eta (1-\zeta),\textbf{p}_{1\rm T}),\quad
 p_2=\frac{M_B}{\sqrt{2}}((1-r^2)(1-\zeta), \eta \zeta,\textbf{p}_{2\rm T}).
\end{eqnarray}

Similar to the situation of the $B$ meson \cite{npb591313},
$B^0_s$ meson could also be treated as a heavy-light system, its wave function in the $b$ space can be
expressed by \cite{prd63054008,prd65014007,epjc28515,ppnp5185}
%{\color{red}  [Refs. needed here]}
\begin{eqnarray}
\Phi_{B_s}(x,b)=\frac{i}{\sqrt{2N_c}}[({ p \hspace{-2.0truemm}/ }_B +M_B)\gamma_5\phi_{B_s}(x,b)],
\end{eqnarray}
where $b$ is the conjugate space coordinate of the transverse momentum $k_{\rm BT}$, and $N_c$ is the color factor.
Here, we only consider one of the dominant Lorentz structures in our calculation.
The distribution amplitude $\phi_{B_s}$ is  adopt the same form as it in Refs.~\cite{prd76074018,prd86-114025}
\begin{eqnarray}
\phi_{B_s}(x,b)=N x^2(1-x)^2\exp[-\frac{x^2M_B^2}{2\omega^2_b}-\frac{\omega^2_bb^2}{2}],
\end{eqnarray}
with  shape parameter $\omega_b=0.50\pm 0.05$ GeV and the normalization constant
$N$ being related to the decay constant $f_{B_s}$ through
\begin{eqnarray}
\int_0^1\phi_{B_s}(x,b=0)d x=\frac{f_{B_s}}{2\sqrt{2N_c}}.
\end{eqnarray}
%$N$ above is  the normalization constant.

For the considered decays, the vector charmonium meson   is
longitudinally polarized. The longitudinal polarized component
of the wave function is defined as~\cite{160208918,epjc75293}
\begin{eqnarray}
\Phi_{\psi}^L=\frac{1}{2\sqrt{N_c}}[m {\epsilon \hspace{-1.5 truemm}/}_L  \phi^L (x,b)+{\epsilon \hspace{-1.5 truemm}/}_L  { p \hspace{-2.0truemm}/ }_3\phi^t (x,b)],
\end{eqnarray}
with the longitudinal polarization vector $\epsilon_L=\frac{M}{\sqrt{2}m}(-r^2,1-\eta, \textbf{0}_{T})$.
For the twist-2 (twist-3) distribution amplitudes $\phi^L (\phi^t)$ of 2S and 3S states,
the same form and parameters are adopted as their in Refs.~\cite{160208918,epjc75293}.

According to \cite{prd91094024}, the S-wave two-pion distribution amplitudes are organized into
\begin{eqnarray}
\phi_{\pi\pi}^S=\frac{1}{2\sqrt{N_c}}[{ p \hspace{-2.0truemm}/ }\Phi_{v \nu=-}^{I=0}(z,\zeta,\omega^2)
+\omega \Phi_{s}^{I=0}(z,\zeta,\omega^2)+\omega ({ n \hspace{-2.0truemm}/ }{ v \hspace{-2.0truemm}/ }-1)\Phi_{t \nu=+}^{I=0}(z,\zeta,\omega^2)],
\end{eqnarray}
where $n=(1,0,\textbf{0}_{T})$  and $v=(0,1,\textbf{0}_{T})$ are two dimensionless vectors.
The asymptotic models for the twist-2 distribution amplitude $\Phi_{v \nu=-}$
and the twist-3 distribution amplitude $\Phi_{s},\Phi_{t \nu=+}$ and relevant  time-like scalar form factor  can be found in \cite{prd91094024}.

Now, we write down the differential   branching ratio for $B^0_s\rightarrow \psi(2S,3S) \pi^+\pi^-$ decays,
%The double differential  $B^0_s\rightarrow \psi \pi^+\pi^-$ branching ratio is
\begin{eqnarray}
\frac{d \mathcal{B}}{d \omega}=\frac{\tau \omega|\overrightarrow{p_1}||\overrightarrow{p_3}|}{32\pi^3M_B^3}|\mathcal{A}|^2,
\end{eqnarray}
where $\tau=1.512\times 10^{-12}s$ is the lifetime of $B^0_s$ meson.
The three-momenta of the pion  in the pion pair center-of-mass system $\overrightarrow{p_1}$ and that
of the charmonium $\overrightarrow{p_3}$ are given by
\begin{eqnarray}
|\overrightarrow{p_1}|=\frac{1}{2}\sqrt{\omega^2-4 m^2_{\pi}},\quad
|\overrightarrow{p_3}|=\frac{1}{2\omega}\sqrt{[M_B^2-(\omega+m)^2][M_B^2-(\omega-m)^2]},
\end{eqnarray}
with $m_{\pi}$ the pion mass.

The decay amplitude $\mathcal{A}$ is written as
\begin{eqnarray}
\mathcal{A}=V^*_{cb} V_{cs}(F^{LL}+M^{LL})-V^*_{tb} V_{ts}(F'^{LL}+F^{LR}+M'^{LL}+M^{SP}),
\end{eqnarray}
with $V_{ij}$ the Cabibbo-Kobayashi-Maskawa (CKM) matrix elements.
The detailed expressions of $F$ (the factorizable emission contributions) and $M$
(the nonfacorizable contributions) are the same as the $B^0_s\rightarrow J/\psi \pi^+\pi^-$ process in the appendix of
Ref.~\cite{prd91094024},
except for the replacement $J/\psi\rightarrow \psi(2S,3S)$.
In this work, we also consider the vertex corrections
to the factorizable amplitudes $F$ at the current known next-to-leading order (NLO) level.
Their effects can be combined in the Wilson coefficients as usual \cite{npb591313,beneke1,beneke2}.
In the NDR scheme we have
\begin{eqnarray}\label{eq:vertex}
a_1&=&C_1+\frac{C_2}{3}+\frac{\alpha_s}{9\pi}C_2[-18-12\text{ln}(\frac{\mu}{m_b})+f_I+g_I(1-r^2)],\nonumber\\
a_2&=&C_3+\frac{C_4}{3}+C_9+\frac{C_{10}}{3}+\frac{\alpha_s}{9\pi}(C_4+C_{10})[-18-12\text{ln}(\frac{\mu}{m_b})+f_I+g_I(1-r^2)],\nonumber\\
a_3&=&C_5+\frac{C_6}{3}+C_7+\frac{C_{8}}{3}+\frac{\alpha_s}{9\pi}(C_6+C_{8})[6+12\text{ln}(\frac{\mu}{m_b})-f_I^h-g_I(1-r^2)].
\end{eqnarray}
The functions $f_I$  and $g_I$ arise from the vertex corrections; they can be found in \cite{prd63074011}.

\section{ results}\label{sec:results}
%In this section, we write down all the values of the parameters required for performing numerical applications.
 The mesons and quark masses (in units of GeV) and the Wolfenstein parameters are taken from the
Particle Data Group~\cite{pdg}
%For numerical calculation,  some input parameters needed in the pQCD calculation which are summarized as follows \cite{pdg}:
\begin{eqnarray}
M_B&=&5.367, \quad m_{\psi(2S)}=3.686,\quad m_{\psi(3S)}=4.040, \quad m_{\pi}=0.14,\quad m_b=4.2,\quad m_c=1.27,\nonumber\\
\lambda &=& 0.22506\pm 0.00050,\quad A=0.811\pm 0.026,\quad \bar{\rho}=0.124^{+0.019}_{-0.018},\quad \bar{\eta}=0.356\pm 0.011.
\end{eqnarray}
The decay constants needed (in units of MeV) are \cite{prl95212001,160208918}
\begin{eqnarray}
f_{B_s}=259\pm 32 ,\quad f_{\psi(2S)}=296^{+3}_{-2} ,\quad f_{\psi(3S)}=187\pm 8.
%f_{B_s}=259\cite{prl95212001},  \quad  f_{\psi(2S)}=296^{+3}_{-2} \cite{160208918}  \quad f_{\psi(3S)}=187\pm 8 \cite{160208918}.
\end{eqnarray}

\begin{figure}[tbp]
%\vspace{-1cm}
\centerline{\epsfxsize=8cm \epsffile{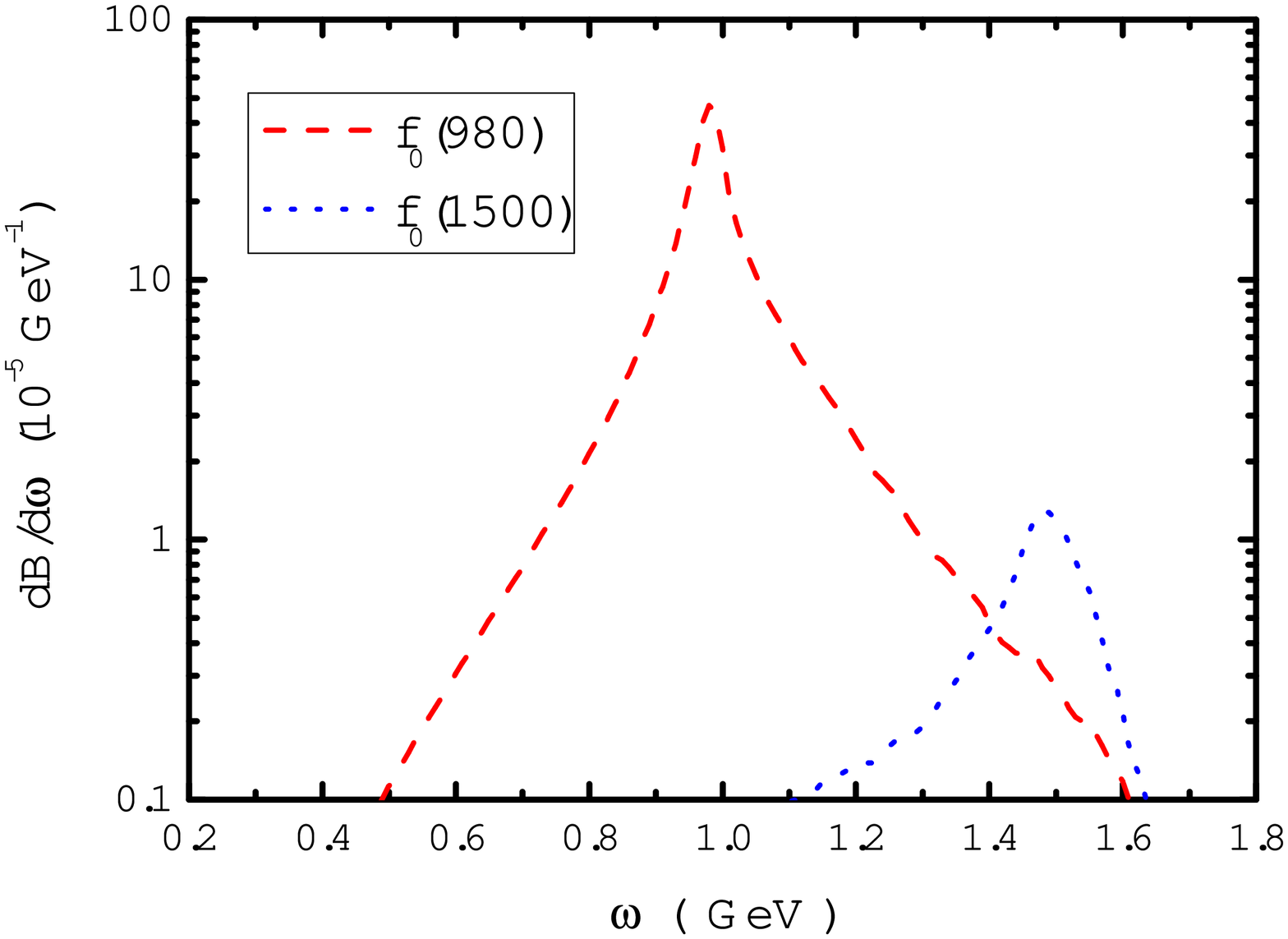}
            \epsfxsize=8cm \epsffile{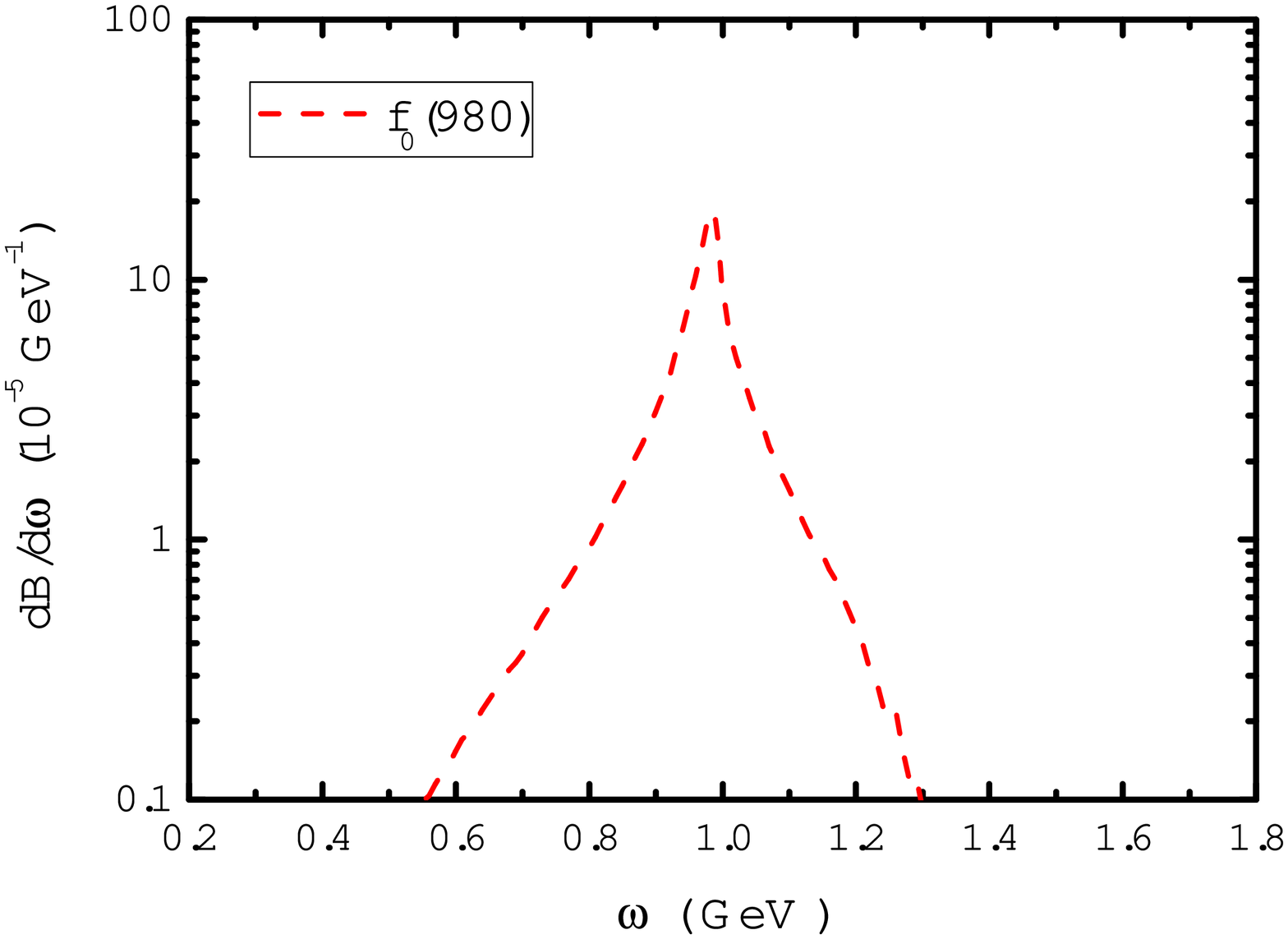}}
\vspace{-0.3cm}
  {\scriptsize\bf (a)\hspace{7.2cm}(b)}
\caption{(a) The $\omega$ dependence of the differential decay rates $d\mathcal{B}/d\omega$ in
(a) $B^0_s\rightarrow \psi(2S)\pi^+\pi^-$ and (b) $B^0_s\rightarrow \psi(3S)\pi^+\pi^-$ decays.}
\label{fig:wave}
\end{figure}

\begin{figure}
\begin{minipage}[b]{0.5\textwidth}
\centering
\includegraphics[height=7.1cm]{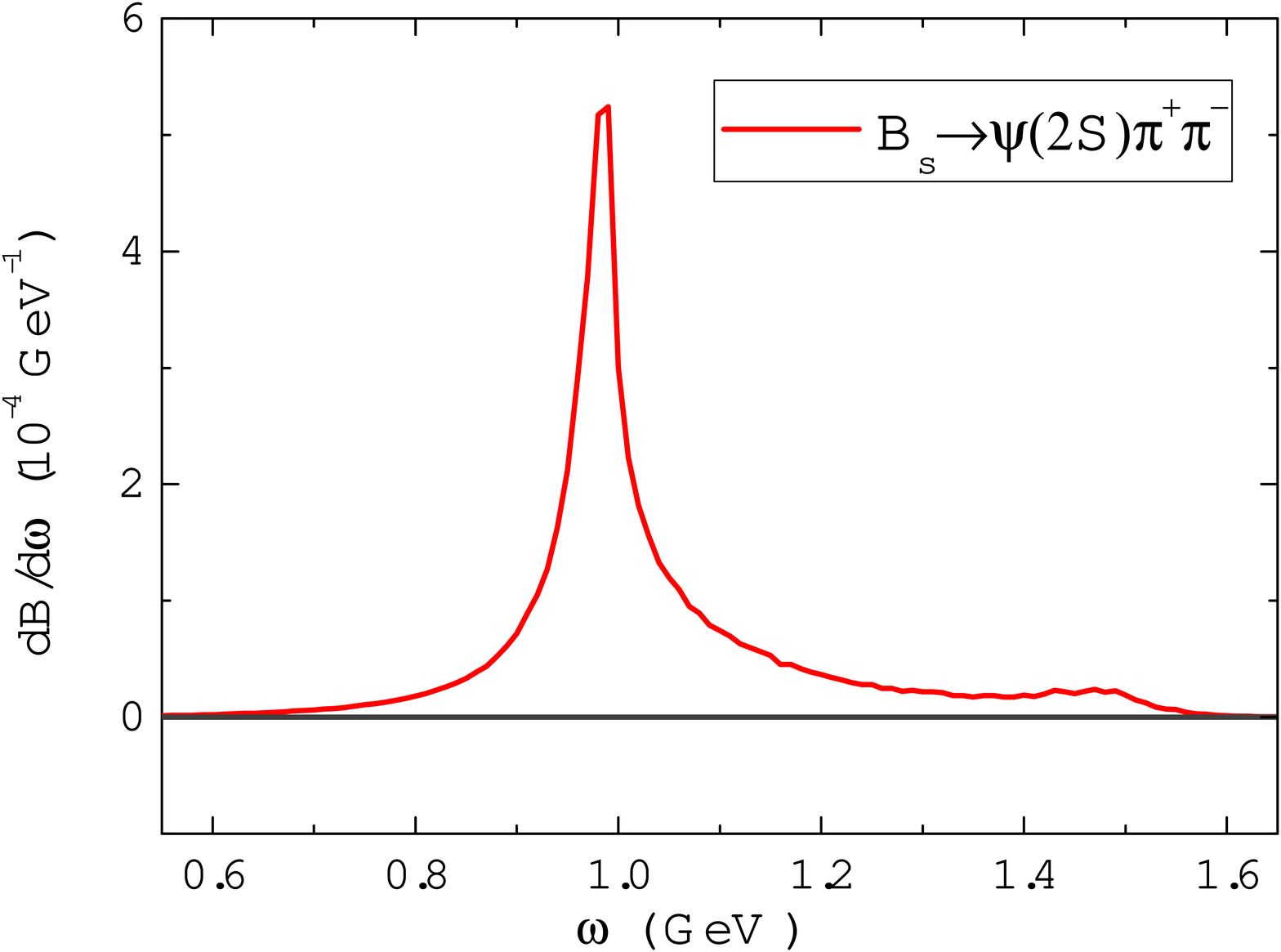}
%\caption{fig1}
%\label{fig:side:a}
\end{minipage}%
\begin{minipage}[b]{0.5\textwidth}
\centering
\includegraphics[height=5.9cm]{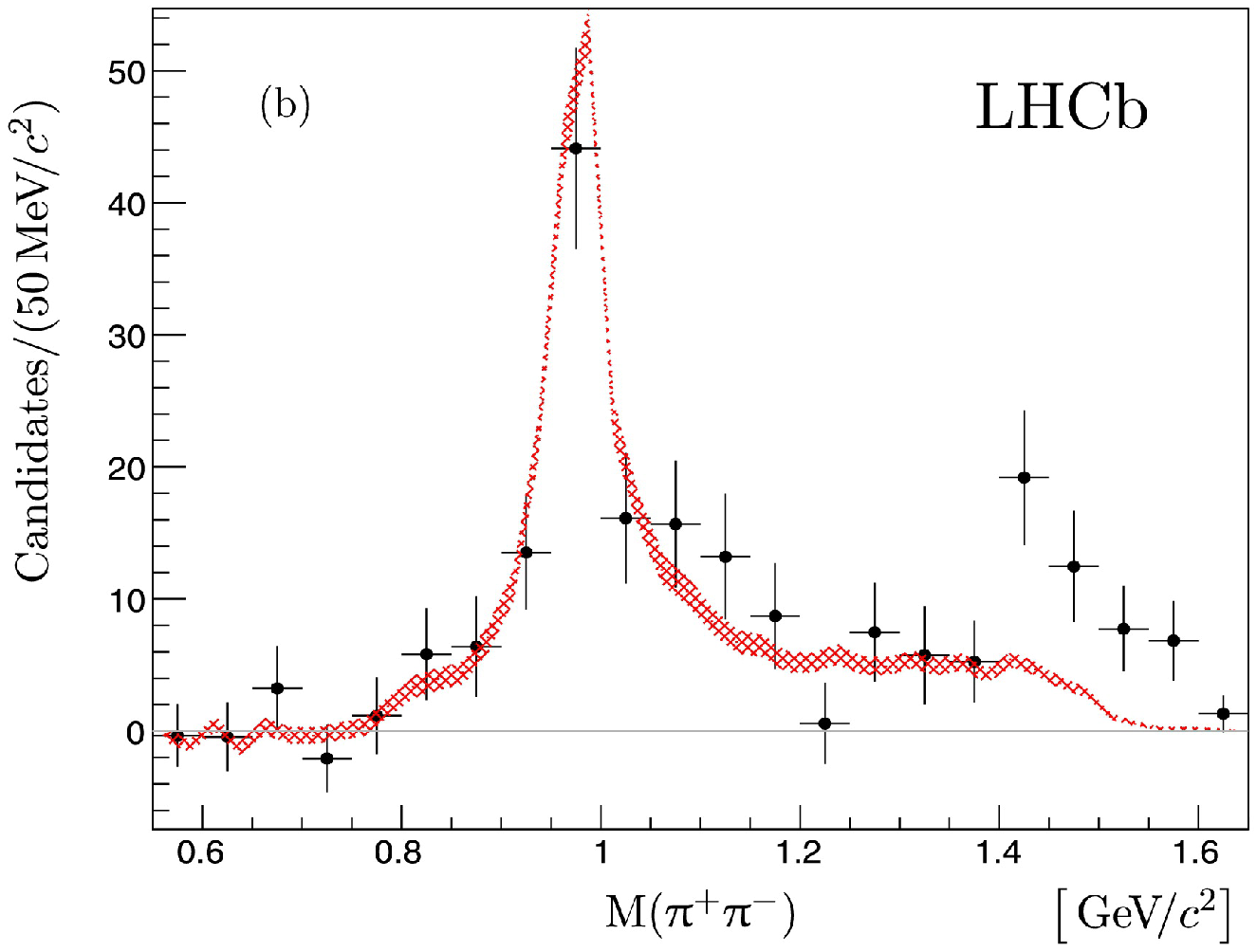}\vspace{0.43cm}
%\caption{fig2}
%\label{fig:side:b}
\end{minipage}
{\scriptsize\bf (a)\hspace{8.cm}(b)}
\caption{ $\pi^+\pi^-$ invariant mass distributions for the $B^0_s\rightarrow \psi(2S)\pi^+\pi^-$ decay.
(a) The S-wave contributions in this work, and (b)  from the LHCb Collaboration \cite{npb871403}.}
 \label{fig:reda}
\end{figure}

By using Eq.~(9), we have the predictions of the branching ratios as
 \begin{eqnarray}\label{eq:branching1}
\mathcal {B}(B^0_s\rightarrow \psi(2S) f_0(980)\rightarrow \psi(2S)\pi^+\pi^- )&=&[5.1^{+1.6}_{-1.2}(\omega_b)^{+1.4}_{-1.2}(f_{B_s})
^{+1.0}_{-0.7}(a_2^{I=0})^{+0.6}_{-0.4}(t)]\times 10^{-5}, \nonumber\\
\mathcal {B}(B^0_s\rightarrow \psi(2S) f_0(1500)\rightarrow \psi(2S)\pi^+\pi^- )&=&[2.3^{+1.3}_{-0.5}(\omega_b)^{+0.7}_{-0.5}(f_{B_s})
^{+0.2}_{-0.1}(a_2^{I=0})^{+0.4}_{-0.1}(t)]\times 10^{-6},
%\mathcal {BR}(B_s\rightarrow  \psi(2S)(\pi^+\pi^- )_S)&=&[6.0^{+1.7}_{-1.4}(\omega_b)^{+1.6}_{-1.4}(f_{B_s})
%^{+0.9}_{-0.9}(a_2^{I=0})^{+0.6}_{-0.5}(t)]\times 10^{-5},
\end{eqnarray}
where the  errors are induced  by the   shape parameter $\omega_b=0.50\pm0.05$ GeV, the decay constant
$f_{B_s}=0.259\pm 0.032$,
%for the $B_s$ meson,
the Gegenbauer coefficient $a_2^{I=0}=0.2\pm 0.2$ for the two-pion system, and the hard scale $t$ which varies from $0.75t$ to $1.25t$, respectively.
 The errors from the uncertainty of the CKM matrix elements and the decay constants of charmonia are very small, and have been neglected.
 It is found that the main uncertainties in our approach come from the $B_s$ meson wave function (the first two errors), which  can reach $30-50 \%$ in size.  The uncertainties caused by the  Gegenbauer coefficient are less than $20 \%$ which is similar to that of $B^0_s \rightarrow J/\psi \pi^+ \pi^-$ \cite{prd91094024}.
The scale-dependent uncertainty is largely reduced due to the inclusion of the next-to-leading order vertex corrections.
%After including the interference between the two  resonances
Summing the two  resonances $f_0(980)$ and $f_0(1500)$ in the  strange  scalar form factors, we have the total
S-wave contribution for  $B^0_s \rightarrow \psi(2S) \pi^+ \pi^-$ decay as
 \begin{eqnarray}\label{eq:branching2}
\mathcal {B}(B^0_s\rightarrow  \psi(2S)(\pi^+\pi^- )_{S-wave})=[6.0^{+1.7}_{-1.4}(\omega_b)^{+1.6}_{-1.4}(f_{B_s})
^{+0.9}_{-0.9}(a_2^{I=0})^{+0.6}_{-0.5}(t)]\times 10^{-5},
\end{eqnarray}
which is compatible with the LHCb data $(7.2\pm 1.2)\times 10^{-5}$ \cite{pdg} when considering its uncertainty.
 %Experimentally, the ratio between the branching ratios of $B_s \rightarrow \psi(2S) \pi^+ \pi^-$ and $B_s \rightarrow J/\psi \pi^+ \pi^-$ is given as
 The measured ratio between the branching ratios of $B^0_s \rightarrow \psi(2S) \pi^+ \pi^-$ and $B^0_s \rightarrow J/\psi \pi^+ \pi^-$ is \cite{npb871403}
 \begin{eqnarray}
\frac{\mathcal {B}(B^0_s\rightarrow \psi(2S) \pi^+ \pi^-)}{\mathcal {B}(B^0_s\rightarrow J/\psi  \pi^+ \pi^-)}
=0.34\pm 0.04 (\text{stat})\pm0.03 (\text{syst})\pm0.01(\mathcal {B}),
\end{eqnarray}
which is consistent with the PQCD prediction $0.37^{+0.25}_{-0.18}$ (uncertainties added in quadrature),  where the  $\mathcal {B}(B^0_s\rightarrow J/\psi \pi^+ \pi^-)$ is read from \cite{prd91094024}.
Because the mass of the  $f_0(1500)$ resonance is beyond the $\pi^+\pi^-$ invariant mass spectra ($2m_{\pi}<\omega<M-m$) in $B^0_s \rightarrow \psi(3S) \pi^+ \pi^-$ decay,
there is only one resonant state $f_0(980)$ contributing to this decay. We have the branching ratio:
 \begin{eqnarray}\label{eq:branching3}
\mathcal {B}(B^0_s\rightarrow \psi(3S) f_0(980)\rightarrow \psi(3S)\pi^+\pi^- )=[1.7^{+0.5}_{-0.3}(\omega_b)^{+0.5}_{-0.4}(f_{B_s})
^{+0.3}_{-0.1}(a_2^{I=0})^{+0.2}_{-0.0}(t)]\times 10^{-5}
\end{eqnarray}

In Fig. \ref{fig:wave}, we plot the differential branching ratio for
$B^0_s\rightarrow \psi(2S,3S)\pi^+\pi^-$ as a function of the $\pi^+\pi^-$ invariant mass~$\omega$.
%The black (solid) curve denotes the total S-wave contribution, while
The  red dashed and blue dotted curves represent the contributions from the  resonances $f_0(980)$ and $f_0(1500)$, respectively.
As expected, the  $f_0(980)$ production is clearly dominant. From the branching ratios in Eqs.~(\ref{eq:branching1}) and  (\ref{eq:branching2}) one could
find that the $f_0(980)$ resonance accounts for $85.0\%$ of the total branching ratio,  the  $f_0(1500)$ resonance  $3.8\%$, and positive interference between the two terms is for $11.2\%$.

 For a more direct comparison with the available experimental data \cite{npb871403}, we also present the S-wave  $\pi^+\pi^-$ invariant mass distributions for the $B^0_s\rightarrow \psi(2S)\pi^+\pi^-$ decay in this work as well as the Fig. 4b in Ref. \cite{npb871403}.
 %as well as the results obtained by LHCb in Fig \ref{fig:reda}.
One observes an appreciable peak  arising form the  $f_0(980)$  resonance and
a less strong, but clearly visible peak for the $f_0(1500)$  in Fig.~3 a.
Comparing with the data,
our distribution below $1.4$ GeV for the resonances agrees quite well,
showing a similar behavior in this region.

\begin{table}\label{tab:br}
\caption{Contributions to the branching ratios ( in unit of $10^{-5}$) of $B^0_s\rightarrow \psi(2S,3S)\pi^+\pi^-$ decays  from the factorizable amplitudes $F$,
from the nonfactorizable amplitudes $M$, and from the interference between $F$ and $M$.}
\label{tab:amp}
\begin{tabular}[t]{lcccc}
\hline\hline
%\toprule[2pt]
Modes                       &FF&MM&FM&Total
 \\ \hline
$B^0_s\rightarrow \psi(2S) f_0(980)\rightarrow \psi(2S)\pi^+\pi^-$ &6.4&0.2&-1.5& 5.1\\
$B^0_s\rightarrow \psi(2S) f_0(1500)\rightarrow \psi(2S)\pi^+\pi^-$ &0.34&0.04&-0.15&0.23\\
$B^0_s\rightarrow\psi(2S)(\pi^+\pi^-)_{\text{S-wave}}$ &7.3&0.3&-1.6&6.0\\
$B^0_s\rightarrow \psi(3S) f_0(980)\rightarrow \psi(3S)\pi^+\pi^-$ &2.0&0.2&-0.5&1.7\\
\hline\hline
\end{tabular}
\end{table}
Different from the fixed kinematics of the two-body decays, the decay amplitudes of the quasi-two-body decays are dependent on the $\pi^+\pi^-$ invariant mass. Therefore,
it is more convenient to compare these contributions to the branching ratios,
whose results are displayed separately in Table \ref{tab:br}, where the
labels $FF$, $MM$, $FM$ correspond to the contribution of
the factorizable, nonfactorizable cases, and the interferences between them, respectively, while the label ``Total" denotes the total contribution. It can be found that the dominant contributions
to the branching ratios  coming from the factorizable topology due to the vertex corrections,
which are enhanced by the Wilson coefficient $C_2$ (see Eq.~(\ref{eq:vertex})).
The interference contributions are of the same order as the factorizable ones with an opposite sign,
which  reflects the importance of nonfactorizable effects  in the color-suppressed processes. This is similar to the case of
the two-body $B$ meson decays into charmonia \cite{prd561615,prd71114008,prd68034004}.
\section{ conclusion}
In this work, the quasi-two-body decays  $B^0_s\rightarrow \psi(2S,3S)\pi^+\pi^-$ have been analyzed
in the PQCD approach, assuming dominance of the
S-wave  resonance states $f_0(980)$ and $f_0(1500)$ in the invariant $\pi^+\pi^-$  mass distributions. Both the factorizable (including the vertex corrections)
and nonfactorizable contributions are  taken into account. %,  and have been evaluated reliably.
We discussed theoretical uncertainties arising from the nonperturbative shape parameters, the decay constant, the Gegenbauer coefficient,  and the scale
dependence. It is found that the main uncertainties of the concerned processes come from the shape parameters and the decay constant of the $B_s$ meson.
The predicted branching ratio  and the invariant mass distributions for $B^0_s\rightarrow \psi(2S)\pi^+\pi^-$ decay are in agreement with the results from LHCb Collaboration.
%compatible with the experimental data within uncertainties.\mathcal {B}(
The decay mode $B^0_s\rightarrow \psi(3S) \pi^+\pi^-$  has not been measured yet,
while the large value of the prediction $ \mathcal {B}=1.7^{+0.8}_{-0.5}\times 10^{-5}$ for it in this work waits   the
the future measurements.

\begin{acknowledgments}
The authors are grateful to Hsiang-nan Li for helpful discussions.
This work was supported in part by the  National Natural Science Foundation of China under
Grants Nos. 11547020, 11605060, and 11547038, in part by the Natural Science Foundation of
Hebei Province  under Grant No. A2014209308, in part by the Program for the Top Young Innovative Talents of Higher Learning Institutions of Hebei Educational Committee under Grant No. BJ2016041, and in part by the Training Foundation of  North China University of Science and Technology  under Grant Nos. GP201520 and  JP201512.
\end{acknowledgments}

\end{document}